\documentclass[nologo,url,11pt,a4paper]{ETHpaper}

\usepackage{amsmath, amsthm, amssymb}           
\DeclareMathOperator*{\sign}{sign}
\DeclareMathOperator*{\arctanh}{arctanh}
\usepackage{graphicx}                  
\usepackage{subfig}                  
\usepackage[numbers,sort&compress]{natbib}

\title{An Agent-Based Model of Collective Emotions \\ in Online
  Communities}

\author{Frank Schweitzer, David Garcia}

\address{Chair of Systems Design, ETH  Zurich, Kreuzplatz 5, 8032 Zurich,
Switzerland}
\titlealternative{An Agent-Based Model of Collective Emotions in Online
  Communities 
\\ \emph{European Physical Journal B} (to appear, 2010). See 
\url{http://www.sg.ethz.ch} for details.
}

\www{\url{http://www.sg.ethz.ch}}

\begin{document}
\maketitle
\begin{center}
  \today
\end{center}

\begin{abstract}
  We develop a agent-based framework to model the emergence of collective
  emotions, which is applied to online communities. Agents individual
  emotions are described by their valence and arousal. Using the concept
  of Brownian agents, these variables change according to a stochastic
  dynamics, which also considers the feedback from online
  communication. Agents generate emotional information, which is stored
  and distributed in a field modeling the online medium. This field
  affects the emotional states of agents in a non-linear manner. We
  derive conditions for the emergence of collective emotions, observable
  in a bimodal valence distribution.  Dependent on a saturated or a
  superlinear feedback between the information field and the agent's
  arousal, we further identify scenarios where collective emotions only
  appear once or in a repeated manner. The analytical results are
  illustrated by agent-based computer simulations. Our framework provides
  testable hypotheses about the emergence of collective emotions, which
  can be verified by data from online communities.
\end{abstract}

\newcommand{\mean}[1]{\left\langle #1 \right\rangle}
\newcommand{\abs}[1]{\left| #1 \right|}

\section{Introduction}
\label{sec:intro}

How do collective phenomena arise from the interaction of many
distributed system elements? This question is certainly at the heart of
statistical physics. Over the last 150 years it has provided a large set
of methodologies applied to physical systems, to infer from the
properties of the elements on the micro level on the systems dynamics on
the macro level. A very similar question is also asked in different other
scientific disciplines. For example, in medicine one wishes to understand
the reaction of the immune system based on the communication and
coordinated action of e.g. B or T cells. In economics one is interested
in the emergence of systemic risk \citep{lorenz} in a financial system
based on the fault of firms or banks clearing their debts to other firms
or banks.

To answer such questions, we need an appropriate description of the
system elements, which are called agents in the following, and their
interactions -- but we also need an appropriate framework to predict from
these ingredients the possible collective dynamics on the systems
level. Without such a framework, we are only left with extensive computer
simulations of multi-agent systems, in which, for given assumptions of
the interactions, we have to probe the entire parameter space, to find
out the conditions for certain collective phenomena.

In this paper, we want to develop such a framework to describe collective
emotions in online communities. There is no commonly accepted definition
of collective emotions yet. According to \citep{Bartal2007}, collective
emotions are shared by large numbers of individuals, in contrast to
group-based emotions that are felt by individuals as a result of their
membership in a certain group or society. The former concept suggests
that group members may share the same emotions for a number of different
reasons, whereas the latter refers to emotions that individuals
experience as a result of identifying with their fellow group members.

Hence, in our paper collective emotions are shared by a larger number of
individuals as a result of both external events and nonlinear coupling
between individuals. Similar to other collective states, also collective
emotions can display new, emergent properties which cannot be traced back
to invididual contributions. Remarkably, the life time of a collective
emotion is usually much larger than the one of an individual emotion. On
the other hand, individual emotions show a different dynamics in the
presence of collective emotions, simply because of the nonlinear feedback
of the emergent collective emotion on the individual one.

In this general manner, collective emotions are not restricted to online
communities. Instead they can emerge in any social context.  The aim of
our research, however, is to understand the dynamics in \emph{online}
communities. The cyberspace does not have emotions, but individuals that
interact online can share emotions.  Here certain conditions of
indiviudal interactions apply that are not present for offline
communities. Online communities react on other time scales (not
necessarily faster, but often with a time shift), they act on different
stimuli (there are hardly seen offline assemblies to share emotions about
a Youtube video), they have different thresholds to express their
emotion, and they do it in a very different manner, namely by writing
in. For all these, we may gather data from online portals, whereas it
becomes very difficult to measure those in offline communities. We remark
that the internet is indeed shaping the phenomena in mind, it is not a
mere interface for monitoring `real' social interaction.  While we agree
that there are certain commonalities (mostly based on social herding and
amplification), there are also substantial differences in communication.
For the very same reason, we argue that real phenomena like mass hystery
can indeed be seen as instances of collective emotions, however, not all
of the modeling implications used in this paper may apply.

Examples of collective emotions in online communities can be observed en
masse on the Internet. One particular example was the large amount of
emotional discussions which followed the death of Michael Jackson, other
examples are the "memes" and heated discussions of anonymous fora like
\url{4chan.org}. They follow a very similar scheme: Users which have
subscribed to social network sites or to blogs or discussion fora, become
enraged or excited about a particular event (like the performance of a
beloved soccer team in a world competition) or a personal (good or bad)
experience. Importantly, these individual feelings are then shared with
other users, i.e. they are \emph{communicated} by means of online media,
most likely by writing a personal statement.  Obviously, users do not
transmit an emotion, instead they communicate a piece of information,
which may trigger an emotional reaction in participants reading
this. Dependent on such an impact, other users may decide to involve
themselves in such an emotional communication, e.g. by sharing the
feeling or opposing to it. Under certain circumstances, we may observe
mutual communication in a small group of users, but there are also
scenarios where many users express their feeling once, in a sequence, or
where many users repeatedly fire the discussions by emotional
statements. These discussions show the existence of emergent collective
states in which the users share their emotions, rather than an aggregate
of the emotions of the community. The discussions do not necessarily have
to be centered around just one feeling. In many cases, we see the
emergence of two collective emotions, a 'positive' and a 'negative' one,
which may coexist or 'fight' each other. These collective states usually
only have a finite life time, i.e. they disappear, but could come back
when triggered by a new event or post.

How do we want to model such collective phenomena? In an agent-based
model, we first need to describe the emotional states of individual
agents, which should be based on insights obtained in psychology. 

First of all, emotions are very different from e.g. \emph{opinions} in
that they are rather short-lived subjective states that decay to the
neutral state very fast (see also the survey study of
\citep{Scherer2004}).  While both emotions and opinions can be influenced
by herding effects, opinion dynamics is often linked to utilities and
preferences, while emotions do not need to follow any particular
optimization space. Further, opinions are often becoming externalized
instances on the collective level (the ``public opinion'' exist outside
the individuals), whereas collective emotions are perceived in a rather
implicite way. We recall that it takes sophisticated algorithms to
extract them from blog entries, etc.

Secondly, emotions are characterized by different dimensions.  An
established theoretical perspective also used in this paper is the
circumplex model \citep{Russell1980} which is based on the two dimensions
of \emph{valence}, indicating whether the pleasure related to an emotion
is positive or negative, and \emph{arousal}, indicating the personal
activity induced by that emotion. However, in the psychology literature
various ways of representing emotions can be found (see
\citep{FeldmanBarret1999,Yik1999} for a review of different dimensional
representations). For the sake of completeness we mention that for
computational models the appraisal theory \citep{scherer01} provides
another promising theoretical perspective: it is based on internal
representations of person-environment relations, which can be modeled by
so-called BDI (belief-desire-intention) agents. While there is a large
body of literature on computational appraisal models
\citep{gratch04,gratch09,melo09} the focus is more on the correct
internal representation of emotions and their cognitive consequences, not
on the explanation of collective phenomena such as described above.

Given that we are able to characterize agent's emotions, how are we able
to \emph{detect} them?  Internal emotional states can be inferred from
physiological signals \citep{Jones2007}, in particular the internal
dynamics and responses of humans in emotional contexts can be measured
under the dimensional representation of \citep{Russell1980}. In online
communities, however, we cannot measure the physiological response of
users directly. Instead, we are left with the problem to infer user's
emotions from the written text pieces they provide in the online media.
Human annotation of internet data was used in \citep{Sobkowicz2010}, but
this largely restricts the amount of data to be processed. Again, in
computational sciences, there are established ways of \emph{sentiment
  mining}, i.e. algorithms to extract the \emph{emotional} content of a
written text and to classify this according to various dimensions.
Different sentiment classification techniques can be combined to improve
results \citep{Prabowo2009} and can be applied to study emotions in the
internet \citep{Thelwall2009}.  Because in this paper we do not provide a
direct comparison of our model results with empirical data, we skip the
detailed discussion of those techniques, keeping in mind that we are
indeed able to obtain e.g. the valences of different users participating
in a blog, over time.
 
Another challenge results from the fact that we need to model the
communication between users in online communities. It is not the emotion
per se of an user what matters, but its expression in a blog entry, a
post etc. This is submitted at a particular time and distributed to the
whole online community, where it is percieved by other users with a very
differerent time delay. While modeling a personalized communication would
need to know the underlying social network, in most online communities a
particular post is available to everyone (who has subscribed)
immediately. This justifies the assumption of a mean-field coupling
between users, i.e. a medium is updated instantaneously and provides the
same information to everyone. Nevertheless, we still have to consider
that for example older posts have less impact on users than more recent
ones and that positive and negative posts may be submitted with different
frequencies. In this paper, we cope with these requirements by
introducing an emotional information field generated by the users, which
stores and distributes this information accordingly. This idea was
already successfully applied in other communication models describing
biological or social systems \citep{schweitzer00b,schweitzer03c}.

Eventually, we need to model the impact of the emotional information on
online users. Unfortunately, the psychological literature does not
provide much insight into this problem. Therefore, we are left with
providing hypotheses about the feedback between the emotional information
and the individual dimensions of arousal and valence.  By proposing a
very general non-linear feedback, for which different special cases are
explored, we are able to derive conditions under which the emergence of
collective emotions can be expected. Different scenarios can be obtained:
either the repeated occurence of collective emotions, or the one-time
collective emotion. These results can be seen as testable hypothesis,
which may be verified either by psychological experiments or by data from
online communities.

In conclusion, in this paper we wish to derive a quite general modeling
approach, to explore the conditions under which collective emotions may
emerge from interacting emotional agents.  Understanding the emergence of
collective emotions certainly has an impact beyond online communities.
They play for example a crucial role in resolving conflicts in societies
\citep{Bartal2007}.  Collective emotions are also important for the
efficiency of working groups \citep{Flache2004}. However, online
communities provide a much better starting point for understanding
collective emotions. First of all, there is a large amount of data
available from these communities. Consequently, large scale emotions have
been already studied for songs, blogs or political comments
\citep{Dodds2009}. In addition to this, some peculiarities of information
exchange between users on the internet are suspected to have an emotional
origin.  For example, the network of posts and comments in various blog
sites \citep{Mitrovic2009} has a strong community structure that could be
created by emotional discussions. The effect of emotions in creating and
reshaping social contacts was also modeled in artificial social networks
\citep{Chmiel2010}.

Secondly, collective emotions are fostered by internet communication
because of (a) the fast information distribution, and (b) the anonymity
of users in the internet, which often seduces people to drift away from
established norms and show a salient private personality. In fact,
empirical studies \citep{Sassenberg2003} which compared the attitude
change in virtual and face to face interactions, have demonstrated that
human behavior and social norms are affected by internet
interaction. Thirdly, the internet is seen as an important factor in
defining present and future societies \citep{DiMaggio2001}. Collective
emotions, such as hate, play an important role in the creation of
collective identites. \citep{Adams2005} provides a systematic study of
collective identity in internet-based hate groups.

\section{An agent-based model of emotions}
\label{sec:agent}

\subsection{The concept of Brownian agents}
\label{sec:brownian}

Our modeling approach is based on the concept of Brownian agents
\cite{schweitzer03c}. It allows to formalize the agent dynamics and to
derive the resulting collective dynamics in close analogy to methods
established in statistical physics. A Brownian agent is described by a
set of state variables $u_{i}^{(k)}$, where the index $i=1,...,N$ refers
to the individual agent $i$, while $k$ indicates the different variables.
These could be either \emph{external} variables that can be observed from
the outside, or \emph{internal degrees of freedom} that can only be
indirectly concluded from observable actions.

Noteworthy, the different (external or internal) state variables can
change in the course of time, either due to influences of the
environment, or due to an internal dynamics. Thus, in a most general way,
we may express the dynamics of the different state variables as follows:
\begin{equation}
  \label{u-ik-d}
  \frac{d\,u_{i}^{(k)}}{dt}=f_{i}^{(k)}+{\cal F}^{\mathrm{stoch}}_{i}
\end{equation}
This formulation reflects the \emph{principle of causality}: any
\emph{effect} such as the temporal change of a variable $u$ has some
\emph{causes} that are listed on the right-hand side. For the concept of
Brownian agents, it is assumed that these causes may be described as a
\emph{superposition} of \emph{deterministic} and \emph{stochastic}
influences, imposed on agent $i$. This distinction is based on Langevins
idea for the description of Brownian motion, which coined the
concept. Hence, we sum up influences which may exist on a microscopic
level, but are not observable on the time and length scale of the
Brownian agent, in a stochastic term ${\cal F}_{i}^{\mathrm{stoch}}$,
while all those influences that can be directly specified on these time
and length scales are summed up in a \emph{deterministic} term
$f_{i}^{(k)}$.  This implies that the ``stochastic'' part does \emph{not}
impose any \emph{directed} influence on the dynamics (which would have
counted as deterministic), but on the other hand, it does not necessarily
mean a white-noise type of stochasticity. Instead, other types such as
colored noise, or multiplicative noise are feasible.  Noteworthy, the
strength of the stochastic influences may also vary for different agents
and may thus depend on local parameters or internal degrees of freedom,
as was already used in different applications \cite{schweitzer03c}.  The
\emph{deterministic} part $f_{i}^{(k)}$ contains all specified influences
that cause changes of the state variable $u_{i}^{(k)}$. This could be
nonlinear interactions with other agents $j\in N$ -- thus $f_{i}^{(k)}$
can be in principle a function of all state variables describing any
agent (including agent $i$). But $f_{i}^{(k)}$ can also describe the
response of an agent to available information, as it will be the case for
cyberemotions.  It should further depend on external conditions -- such
as forces resulting from external influences (most notably information
from mass media).  Eventually, $f_{i}^{(k)}$ may reflect an (external or
internal) \emph{eigendynamics} -- in the considered case a relaxation of
the excitited emotional state of an agent (caused by saturation or
exhaustion).  In order to set up a multiagent system (MAS) we need to
specify the relevant state variables $u_{i}^{(k)}$ and the dynamics of
their change, i.e.  $f_{i}^{(k)}$, which means also to specify the
interaction between the agents. We emphasize that the dynamics of the MAS
is specified on the level of the individual agent, not on a macroscopic
level, so the collective dynamics shall emerge from the interactions of
many agents.

\subsection{Emotional states}
\label{sec:emotions}

To quantify the emotional dynamics of an agent, we consider the following
continuous variables, \emph{valence}, $v_{i}(t)$, and \emph{arousal},
$a_{i}(t)$. Both define a two-dimensional
plane 
for the classification of emotions.  Valence (x-axis) measures whether an
emotion is positive or negative, arousal (y-axis) measures the degree of
personal activity induced by that emotion. Hence, an emotional state is
defined by $e_{i}(t)=\{v_{i}(t),a_{i}(t)\}$. For example, `astonished' is
an emotional state with both positive valence and arousal, 'satisfied'
has a positive valence, but a negative arousal, 'depressed' has both a
negative valence and arousal, and 'annoyed' has a negative valence and a
positive arousal.

We note that both valence and arousal describe \emph{internal} variables,
describing a dynamics inside the agent, which may be only indirectly
observable, for example through physiological measurements.

Without any internal or external excitation, there should be no positive
or negative emotion, so we assume that in the course of time both valence
and arousal relax into an equilibrium state, $e_{i}(t)\to 0$, which
implies $v_{i}(t)\to 0$, $a_{i}(t)\to 0$. Hence, in accordance with
eqn. (\ref{u-ik-d}) we specify the dynamics of the Brownian agent as
follows:
\begin{eqnarray}
  \label{eq:v-basic}
  \dot{v_i} &=& - \gamma_{vi}\, v_i(t) + \mathcal{F}_v +
  A_{vi}\;\xi_v(t)  \\   \label{eq:a-basic}
  \dot{a_i} &=& - \gamma_{ai} \, a_i(t) + \mathcal{F}_a + 
  A_{ai}\,\xi_a(t)
\end{eqnarray}
The first term in each equation describes the relaxation into an
equilibrium state as an exponential decay of both valence and arousal, if
no excitation is given. $\gamma_{vi}$, $\gamma_{ai}$ define the time
scales for this relaxation, which are different for valence and arousal
and further may vary across individual agents. The second and third term
in the equations above describe influences which may induce an emotional
state. These can be stochastic influences, expressed by the third term,
where $\xi_{v}(t)$, $\xi_{a}(t)$ are random numbers drawn from a given
distribution of stochastic shocks, with the mean of zero
$\mean{\xi(t)}=0$ and no temporal correlations between subsequent events
$\mean{\xi(t)\xi(t')}=\delta(t-t')$. $ A_{vi}$, $ A_{ai}$ denote the
strength of these stochastic influences which may again vary across
agents. The two functions $\mathcal{F}_v$, $\mathcal{F}_a$ describe
deterministic influences which cause the emotional state. They very much
depend on the specific assumptions applicable to collective
cyberemotions, in particular the agents' interaction, access to
information, response to the media, but can depend also on internal
variables such as \emph{empathy}, i.e. the ability to share the feelings
of other agents, or responsiveness to available information. Most of all,
these functions should also reflect a dependence on the emotional state
itself, i.e. agents already in a specific mood may be more affected by
particular emotions of others. Before we specify these functions in
detail, we need to extend the agent description.

\subsection{Emotional actions}
\label{sec:actions}

The dynamics of eqs. (\ref{eq:v-basic}), (\ref{eq:a-basic}) already
define a stationary state $e_{i}(t)\to 0$, given that the deterministic
and stochastic influences become negligible. On the other hand, there
should be an excited emotional state of the agent if these influences are
large, e.g. if information with a large emotional content becomes
available to the agent. Per se, this state is not observable from the
outside unless the agent takes any action that \emph{communicates} that
emotional state, for example by posting in a blog, etc.  Consequently, we
assume that the agent expresses its \emph{valence}, i.e. the good or bad
feeling, \emph{if} its \emph{arousal}, i.e. the action induced by the
emotion, exceeds a certain individual threshold, $\tau_{i}$:
\begin{equation}
  \label{s-i}
  s_i(t+\Delta t) = \sign(v_i(t))\, \Theta[a_{i}(t) - \tau_i]
\end{equation}
Here $\Theta[x]$ is the Heavyside function which is one only if $x\geq 0$
and zero otherwise. If $\Theta[x]=1$, we make the simplifying assumption
that the agent does not communicate all details about his feelings
(i.e. the value of $v_{i}$) because perfect emotional information cannot
be communicated. Instead, the agent communicates only if it is a good or
bad feeling, i.e. the sign of $v_{i}$, -1 or +1, which is defined as
$r_{i}(t)=\sign(v_i(t))$ in the following (Note: the model specified here
is not really changed if indeed the $v_{i}$ is communicated, but the
analytical investigations become more involved). This coarse-grained
description of the valence only enters the communication process, while
valences are still distributed across agents.

Eqn. (\ref{s-i}) further reflects the assumption that the agent does not
immediately express its feelings if the arousal hits the threshold at
time $t$, but probably with a certain delay $\Delta t$, which may be
caused by the fact that the agent has no immediate access to some
communication media (computers in the case of cyberemotions) or other
things to do. More important feelings should be communicated with a
shorter delay. It should vary as well across agents. In accordance with
investigations of waiting time distributions in performing human
activities (e.g. answering emails), we may assume that $\Delta t$ can be
random drawn from a power-law distribution $P(\Delta t)\propto \Delta
t^{-\alpha}$, where $\alpha$ should be empirically determined. Note that
the dynamics of the external state variable $s_{i}(t)$ differ from the
form given in eqn. (\ref{u-ik-d}) in that the stochastic influences are
not additive, but implicitely present because of the stochastic dynamics
for $v_{i}(t)$ (determining the \emph{sign} of the expression),
$a_{i}(t)$ (determining the \emph{time} of the expression) and $\Delta t$
(determining the \emph{delay} of the expression).

Based on eqn. (\ref{s-i}), we can define the number of emotional
expressions at a given time $t$ as
\begin{equation}
  \label{eq:ns}
  N_{s}(t)=\sum_{i}  \Theta[a_{i}(t) - \tau_i]
\end{equation}
Assuming continuous time, the \emph{average} number of expressions per
time interval then results from 
\begin{equation}
  \label{eq:ns-average}
  n_{s}=\frac{1}{t_{\mathrm{end}}}\int_{0}^{t_{\mathrm{end}}}N_{s}(t) dt
\end{equation}
We may calculate this quantity from analytical approximations in
Sect. \ref{sec:expression} and from computer simulations in
Sect. \ref{sec:simulation}.

\subsection{Communicating emotions}
\label{sec:communication}

By now we have described the (internal) emotional dynamics of an agent
that leads to a certain (externally visible) expression of an emotional
state. In order to describe cyberemotions as collective emotions, we now
need to specify how this emotional expression is communicated to other
agents. In accordance with previous investigations \cite{schweitzer00b}
we assume that every positive or negative expression is stored in a
\emph{communication field} $h_{+}(t)$ or $h_{-}(t)$ dependent on its
value. $h_{\pm}(t)$ represent the communication media available for the
storage and distribution of emotional statements, for example blogs,
forums, etc. and simply measure the 'amount' of positive or negative
feelings available at a given time. For the dynamics of the field, we
propose the following equation:
\begin{equation}
  \label{h}
  \dot{h}_{\pm} = - \gamma_\pm h_{\pm}(t) + 
  s N_{\pm}(t) + I_{\pm}(t)
\end{equation}
Each agent contribution $s_{i}(t)$ increases the respective field $h_{+}$
or $h_{-}$ by a fixed amount $s$ at the time of expression, which
represents the impact of the information created by the agent in the
information field, as a time scale parameter. $N_{\pm}(t)$ is the total
number of agents contributing positive or negative statements at a given
time $t$, i.e. all the agents with $s_{i}(t) = 1$ and with $s_{i}(t) =
-1$ respectively.

The relevance of contributions fade out over time as
e.g. agents become less affected by old blog entries. This is covered by
an exponential decay of the available information with the time scales
$\gamma_{\pm}$. Eventually, in addition to the agent contributions,
positive or negative emotional content from the news may add to the
communication field, which is covered by an agent-independent term
$I_{\pm}(t)$, which can be modeled for example by a stochastic input.

\begin{figure}[h]
  \centering \includegraphics[width=0.48\textwidth]{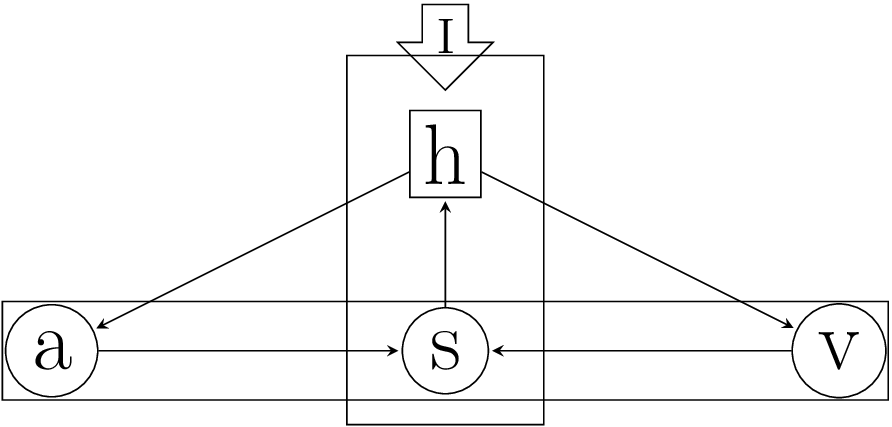}
  \caption{Causation among the components of the model.}
  \label{fig:general-schema}
\end{figure}

The main feedback loops of this framework are sketched in Fig.
\ref{fig:general-schema}, where we can distinguish between two layers: an
internal layer describing the agent (shown horizontally) and an external
layer describing the communication process (shown vertically). In the
internal layer, the arousal $a$ and the valence $v$ of an agent determine
its emotional expression $s$, which reaches the external layer by
contributing to the communication field $h$. The latter one has its
independent dynamics and can, in addition to contributions from other
agents, also consider input from external sources, $I$. The causality is
closed by considering that both valence and arousal of an agent are
affected by the communication field.

In order to complete the model, we need to specify how the available
information affects the emotional states of the individual agents, which
is covered in the functions $\mathcal{F}_{v}$ and $\mathcal{F}_{a}$.

\subsection{Emotional feedback}
\label{sec:feedback}

Because we are interested in the outbreak of \emph{collective} emotion,
we do not assume the latter as the simple superposition (or addition) of
individual emotional states. On the contrary, we assume that an emotional
state of one agent, if it is expressed and communicated to other agents,
may affect the emotional state of these agents either directly or
indirectly. Regarding this effect we are left with hypotheses at the
moment. These could be tested in computer simulations to investigate
their impact on the possible emergence of a collective emotion -- as it
is done in the following. But there should be also the possibility to
empirically test how individuals are affected by different emotional
content, as discussed e.g. in \cite{gianotti2008first}.

With respect to the valence, i.e. the good or bad feeling, we have to
take into account that there are two different kind of emotions in the
system, positive ones represented by $h_{+}(t)$ and negative ones
represented by $h_{-}(t)$. Dependent on its own emotional state, an agent
may be affected by these information in a different way.  If we for
example assume that agents with negative (positive) valence mostly
respond to negative (positive) emotional content, we have to specify:
\begin{equation}
  \label{fv-own}
  \mathcal{F}_{v} \propto \frac{r_{i}}{2}\left\{(1+r_{i}) f[h_{+}(t)] - (1-r_{i})
    f[h_{-}(t)] \right\}
\end{equation}
where $r_{i}(t)=\sign(v_{i}(t))$ and $f(h_{\pm}(t))$ are some functions
depending either on $h_{+}$ or on $h_{-}$ only. Eqn. (\ref{fv-own}) then
results in $\mathcal{F}_{v} \propto f[h_{-}(t)]$ if the agents have a
positive valence ($r_{i}(t)=+1$), and in $\mathcal{F}_{v} \propto
f[h_{+}(t)]$ in the case of negative valence ($r_{i}(t)=-1$).

If, on the other hand, there is evidence that agents, independent of
their valence, always pay attention to the prevalence of positive or
negative emotional content, we may assume:
\begin{equation}
  \label{fv-diff}
  \mathcal{F}_{v} \propto  g[h_{+}(t)-h_{-}(t)]
\end{equation}
where $g$ is some function of the difference between the two information
available. Other combinations, for example agents with positive
(negative) valence pay more attention to negative (positive) emotional
content, can be tested as well.  Some studies in psycho-physiology
\citep{bradley2009} provide initial results for heterogeneous emotional
attention processes related to valence and arousal.

In the following, we may assume the case of eqn. (\ref{fv-own}), i.e. the
valence increases with the respective information perceived by the agent.
The impact, however, should depend also on the emotional state of the
agents in a nonlinear manner.  I.e. if an agent is happy (sad), it may
become happier (more sad) if receiving information about happy (sad)
agents or events, in a nonlinar manner, expressed in the general form:
\begin{equation}
  \label{fv}
  f[h_{\pm}(t),v_i(t)] = h_{\pm}(t) \sum_{k=0}^n b_k v^{k}(t) 
\end{equation}
Here, it is assumed that the coefficients $b_{k}$ are the same for
positive and negative valences, which of course can be extended toward
different coefficients.

\subsection{Arousal and threshold}
\label{sec:arousal}

While the valence expresses the positivity or negativity of the emotion,
the arousal measures the degree in which the emotion encourages or
disencourages activity. Only the latter is important for communicating
the emotional content, which happens if a threshold $\tau_{i}$ of
the arousal is reached. Certainly, expressing the emotion should have
some impact of the arousal, e.g. it is legitimate to assume that the
arousal is lowered because of this action, or set back to the initial
state in the most simple case. That means we should split the dynamics
for the arousal into two parts, one applying before the threshold is
reached, the other one when it is reached. For this, we redefine the
arousal dynamics for $a_{i}(t)$ given in eqn. (\ref{eq:a-basic}) as the
subthreshold dynamics $\dot{\bar{a}}_{i}(t)$ and set:
\begin{equation}
  \label{aorusal-new} 
  \dot{a}_i = \dot{\bar{a}}_{i}(t) \, \Theta[ \tau_i - a_i(t)] 
  - a_i(t)\, \Theta[a_i(t) - \tau_i]
\end{equation}
As long as $x=\tau_i - a_{i}(t)>0 $, $\Theta[x]=1$ and the arousal
dynamics is given by $\dot{\bar{a}}_{i}(t)$, eqn. (\ref{eq:a-basic}),
because $\Theta[-x]=0$.  However, after the threshold is reached, $x\geq
0$, $\Theta[x]=0$ and $\Theta[-x]=1$, i.e. the arousal is reset to zero.

It remains to specify the function $\mathcal{F}_{a}$ for the subthreshold
arousal dynamics. Since arousal measures an activity level, it would be
reasonable to assume that agents respond to the sum of both positive and
negative emotional content in a way that also depends on their own
arousal in a nonlinear manner, regardless of the valence dimension. So,
similar as for the valence, we may propose the nonlinear dependence:
\begin{equation}
  \label{fa}
  \mathcal{F}_{a} \propto  [h_{+}(t)+h_{-}(t)] \sum_{k=0}^n d_k a^{k}(t) 
\end{equation}
Differently from the above assumption, we may argue that agents pay
attention to the information only as long as their arousal is positive
because negative arousals are associated with states of inactivities
(tired, sleepy, depressed, bored). In this case, it is reasonable to
assume e.g. that the impact of information increases linearly with the
activity level:
\begin{equation}
  \label{fa-used}
  \mathcal{F}_{a} \propto  [h_{+}(t)+h_{-}(t)] \;a(t) \;\Theta[a(t)] 
\end{equation}
To conclude the above description, we have set out a model where agents
emotions are characterized by two variables, valence and arousal. These
variables can be psychologically justified and most likely proxied
empirically. The combination of these defines what kind of emotional
content the agent expresses as an observable output. Again, this output
is measurable and can be analysed. The way the emotional content is
stored and distributed to other agents is explicitely modeled as part of
a communication dynamics, which can be adjusted to specific practical
situations.

\section{Emergence of collective emotions}
\label{sec:sim}

\subsection{Valence dynamics}
\label{sec:valence-sim}

In our model, a collective emotional state can only emerge if a
sufficient number of agents expresses their individual valences, which in
turn depends on their arousal. The latter one gets above a critical
threshold only if there is sufficient the emotional information
$h_{+}(t)$, $h_{-}(t)$ available. However, this information is generated
only by the agents.  Hence, there is a circular causality between
$h(t)=h_{+}(t)+h_{-}(t)$ and $a_{i}(t)$.

In order to get a first insight into the dynamics, let us assume that
there exist two different regimes: (i) a 'silent' regime where no
sufficient emotional information is available, i.e. $h(t)\to 0$, and (ii)
an 'excited' regime, where $h(t)$ becomes large enough to affect enough
agents. To simplify the study, we also assume that each agent is mostly
affected by the information that corresponds to its valence state, as
given by eqn. (\ref{fv-own}).

Then, neglecting any sort of random influences, the dynamics of the
valence is expressed by:
\begin{equation}
  \label{cubic}
  \dot{v}= - \gamma_v v(t) +
  h_{\pm}(t)\left\{b_{0}+b_{1}v(t)+b_{2}v^{2}(t)+b_{3}v^{3}(t) +
    ...\right\}
\end{equation}
The stationary solutions for the valence then follow from the cubic
equation:
\begin{equation}
  \label{eq:cubic2}
  v^{3}+v^{2}\{b_{2}/b_{3}\}+v\{(b_{1}-\gamma_{v}/h_{\pm})/b_{3}\}+ \{b_{0}/b_{3}\}=0
\end{equation}
This allows to discuss the following cases:
\begin{itemize}
\item In order to allow for a solution $v\to 0$ as requested, $b_{0}$
  should tend to zero as well, so we use $b_{0}=0$ here. This leads to
  \begin{equation}
    \label{eq:cubic-b0}
    v\left[v^{2}+v\{b_{2}/b_{3}\}+\{(b_{1}-\gamma_{v}/h_{\pm})/b_{3}\}\right]=0
  \end{equation}

\item If positive and negative valences are treated as 'equal', there
  should be no ab initio bias towards one of them, which implies
  $b_{2}=0$. This gives, in addition to $v=0$, the following two
  solutions:
  \begin{equation}
    \label{eq:cubic-b2}
    v^{2}= \frac{b_{1}-\gamma_{v}/h_{\pm}}{b_{3}}
  \end{equation}
  These two solutions become real only if $b_{1}> \gamma_{v}/h_{\pm}$. In
  this case, we have two equilibrium states for the valences which are
  symmetrical wrt to zero. Otherwise, $v=0$ is the only possible
  real solution.
\end{itemize}
So, dependent on the value of the information field $h_{\pm}$ we can
expect the two regimes: (i) the \emph{silent regime} with $h_{\pm}\to 0$
and $v=0$, and (ii) the \emph{excited regime} with the emergence of two
different emotions, each of them centered around $\pm b_{1}/b_{3}$
(provided the field is large enough). We note that these solutions are
symmetrical, which can be changed by considering (a) a bias in the
response ($b_{2}\neq 0$) or (b) differences in the two informations
$h_{\pm}$ (e.g. via different decay rates). It remains to be discussed
whether a \emph{coexistence} between the two collective emotional states
is possible or the \emph{prevalence} of one of them results. This leads
us to the question of path dependence and emotional feedback of section
\ref{sec:arousal-sim}.

\begin{figure}[htbp]
  \centering
    \includegraphics[width=0.48\textwidth]{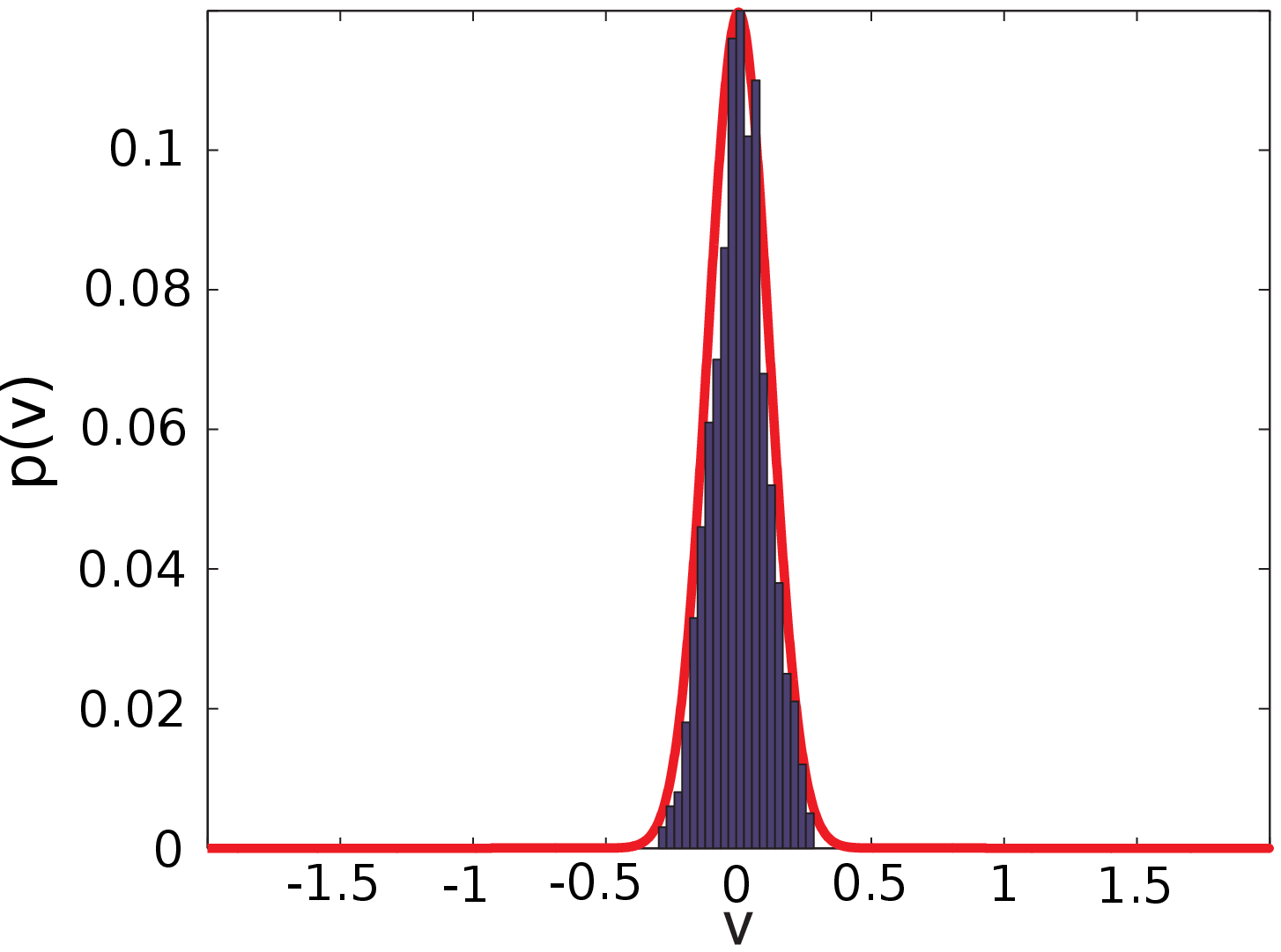} \hfill
    \includegraphics[width=0.48\textwidth]{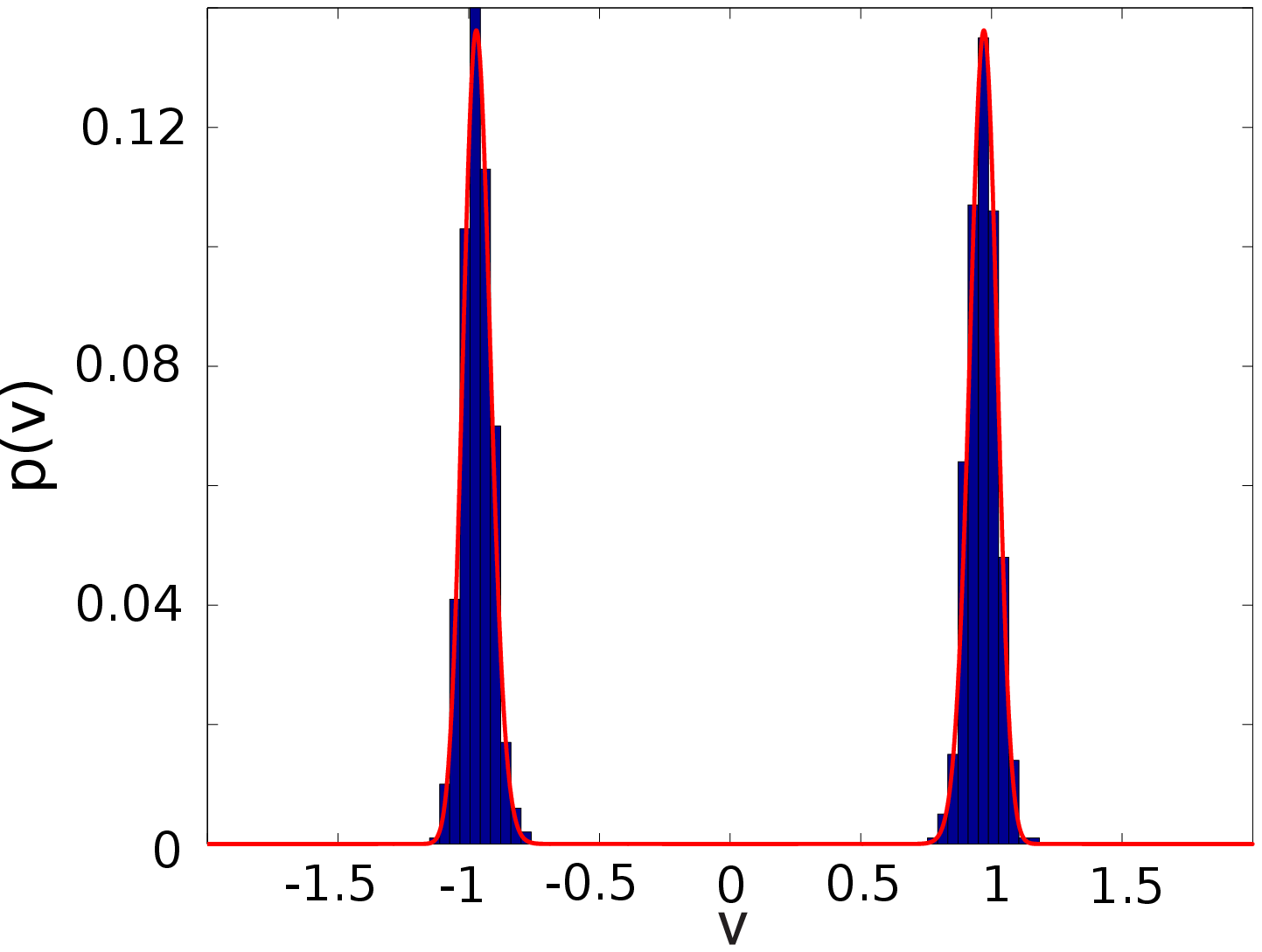}
    \caption{Analytical prediction of the valence distribution from
      eqn. \ref{eq:sol-FP-valence} and histogram of valences for $h=0.25$
      (left) and $h = 17.5$ (right) with parameters $N=1000$,
      $A_{v}=0.15$, $\gamma_{v}=0.8$, $b_{1}=1$ and $b_{3}=-1$.}
  \label{fig:valence-dist}
\end{figure}

If, in addition to the deterministic dynamics specified above, we further
consider stochastic influences as specified in the Langevin dynamics,
eqn.  (\ref{eq:v-basic}), we can write up a dynamics for the valence
distribution $p(v,t)$. Using eqs. (\ref{eq:v-basic}), (\ref{eq:cubic2}),
this is given by the following Fokker-Planck equation:
\begin{equation}
  \label{eq:FP-valence}
  \partial_t p(v,t) = - \partial_v \left[(b_1h_{\pm}-\gamma_v)v - b_3h_{\pm} v^3\right]\;
  p(v,t) + \frac{A_v^2}{2}\; \partial_{v}^{2}p(v,t)
\end{equation}
The stationary solution of the Fokker-Planck equation, $\partial_t
p(v,t) = 0$, reads as:
\begin{equation}
  \label{eq:sol-FP-valence}
  p(v) = \frac{1}{\mathcal{N}_{v}} \exp{\left\{ \frac{v^{2}(b_{1}h_{\pm} -
        \gamma_{v}) - v^{4}(b_{3}h_{\pm}/2)}{A_{v}^{2}} \right\}}
\end{equation}
$\mathcal{N}_v$ is the normalization constant resulting from
$\int_{-\infty}^{\infty}p(v)dv = 1$. In accordance with the discussion
above, the stationary valence distribution $p(v)$ is \emph{unimodal} with
the maximum at $v=0$ if $b_{1}< \gamma_{v}/h_{\pm}$ and \emph{bimodal}
with the maxima given by eqn. (\ref{eq:cubic-b2}) if $b_{1}>
\gamma_{v}/h_{\pm}$. In both cases, the variance of the distribution is
determined by the strength of the stochastic force, $A_{v}^{2}$. This is
shown in Fig. \ref{fig:valence-dist} for two different values of $h$.
The histograms result from computer simulations of the stochastic valence
dynamics of 1000 agents for a given $h$, whereas the solid curves are
given by the analytical solution of eqn. (\ref{eq:sol-FP-valence}).

To conclude, this analysis has provided us with conditions regarding the
influence of emotional information and the response to it, which may lead
to the emergence of a collective emotional state. These conditions can be
seen as testable hypotheses about the feedback between emotional
information and individuals. If they hold true, we are able to predict
the valence of a collective emotional state - which can then be compared
to empirical findings. Deviations from these findings, on the other hand,
allow us to successively refine the modeling assuptions made. Thus, the
framework provided is a useful step toward a thorough understanding of
collective emotions.

\subsection{Arousal dynamics}
\label{sec:arousal-sim}

In the previous section, we have detected an excited regime with a
non-trivial valence, based on the assumption that the emotional
information $h(t)$ is large enough to affect the agents. The generation
of such information, however, depends on the arousal
dynamics. Specifically, in our model the arousal needs to reach a certain
threshold $\tau_{i}$ at which the agent expresses its emotion.  In
Sect. \ref{sec:arousal}, we have already introduced an arousal dynamics,
eqn. (\ref{aorusal-new}), which distinguishes between a subthreshold
dynamics, eqn. (\ref{eq:a-basic}), and a dynamics at the threshold.  By
using the nonlinear assumption of eqn. (\ref{fa}) up to second order and
neglecting stochastic influences for the moment, we get for the
subthreshold regime (omitting the index $i$ for the moment):
\begin{equation}
  \label{eq:arousal-square}
  \dot{a}= - \gamma_a a(t) +
  h(t)\left\{d_{0}+d_{1}a(t)+d_{2}a^{2}(t)+ 
    ...\right\}
\end{equation}
where $h(t)=h_{+}(t)+h_{-}(t)$. The stationary solutions follow from:
\begin{equation}
  \label{eq:square2}
  a^{2}+a\{(d_{1}-\gamma_{a}/h)/d_{2}\}+ \{d_{0}/d_{2}\}=0
\end{equation}
which allows to discuss the following cases. If we consider only the
constant influence of $h$, i.e. $d_{0}\neq 0$, $d_{1}=d_{2}=0$, or a
linear increase with $a$, i.e. $d_{0}\neq 0$, $d_{1}\neq 0$, $d_{2}=0$,
we arrive at only \emph{one} stationary solution for the arousal, which
depends on $h$:
\begin{equation}
  \label{eq:a-lin}
  a(h)=\frac{h d_{0}}{\gamma_{a}-d_{1}h}
\end{equation}
It means that the agents tend to be always in an 'excited' regime, the
level of which is determined by $h$. If it happens that the arousal
reaches a value above the threshold, $a(h)\geq \tau$, then $a(t)$
is arbitrarily set back to zero and then starts to reach $a(h)$, again.
The proposed 'silent' regime would then be reached only if $h\to 0$.

\begin{figure}[htbp]
  \centering
    \includegraphics[width=0.48\textwidth]{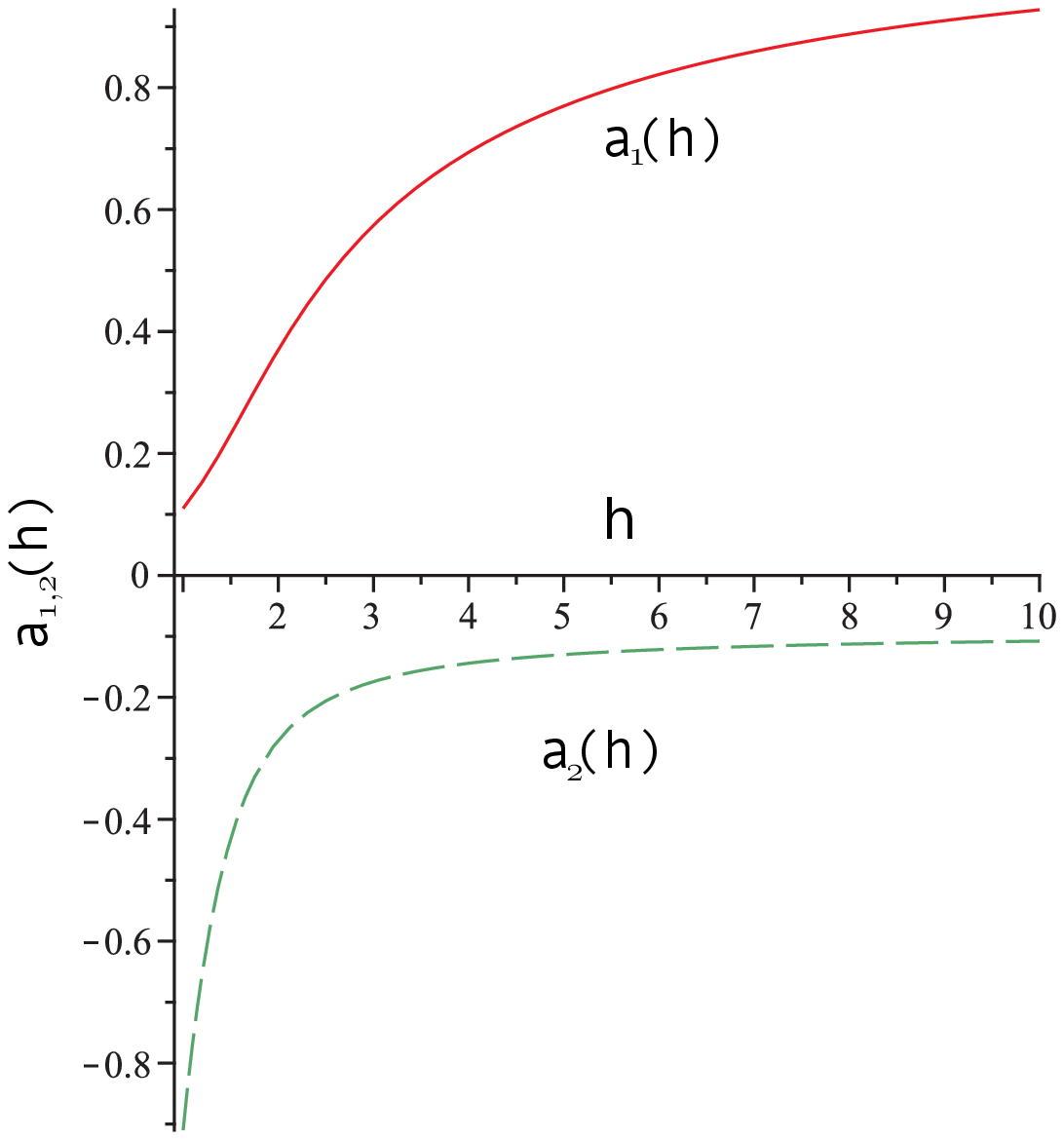}
  \hfill 
    \includegraphics[width=0.48\textwidth]{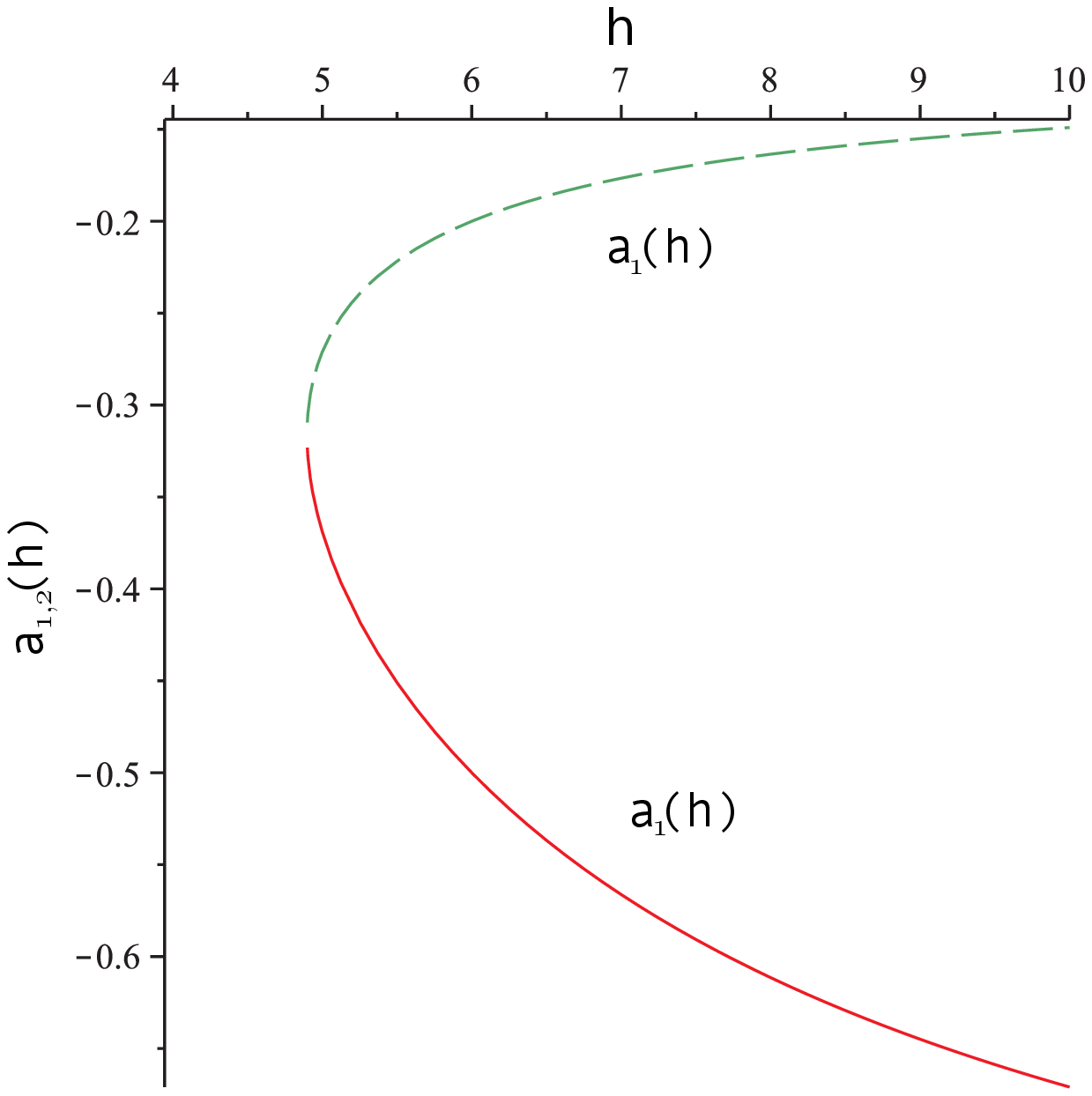}
  \caption{$a_{1,2}(h)$ for the parameter set \ref{eq:parameters}. There is no
    bifurcation present when $d_2=-0.5$ (left) but it appears when
    $d_2=0.5$ (right)}
  \label{fig:arousalSolutions}
\end{figure}

In order to allow for a dynamics where agents can stay at low values of
the arousal even if $h$ is large, we have to consider a non-linear
influence of the emotional information $h$, i.e. $d_{0}\neq 0$,
$d_{1}\neq 0$, $d_{2}\neq 0$ in the most simple case.
Eqn. (\ref{eq:square2}) then has two solutions
\begin{equation}
  \label{eq:a-square}
  a_{1,2}(h)= \frac{1}{2}\left(\frac{\gamma_{a}/h-d_{1}}{d_{2}}\right)
  \pm \sqrt{ \frac{1}{4}\left(\frac{d_{1}-\gamma_{a}/h}{d_{2}}\right)^{2}
    -\frac{d_{0}}{d_{2}}}
\end{equation}  
which are real only if
\begin{equation}
  \left(\frac{d_{1}-\gamma_{a}/h}{2\,d_{2}}\right)^{2}
  > \frac{d_{0}}{d_{2}}
  \label{eq:restr}
\end{equation}
From this restriction, we can infer some important conditions on the
arousal dynamics. Provided $d_{0}>0$, inequality (\ref{eq:restr}) is
always fulfilled if $d_{2} < 0$ which, for a given $h$, implies a
saturation in the feedback of the arousal on the arousal dynamics. Then
we always have two real stationary solutions for the arousal, a positive
and a negative one, shown in Fig. \ref{fig:arousalSolutions}. While the
positive solution is stable for all values of $h$, the negative one is
always unstable, as verified by the second derivative. This allows to
infer the following dynamics for agents expressing their emotions: For
agents starting with a small positive or negative arousal, $a>a_{2}(h)$,
$a(t)$ may grow in time up to the stationary value $a_{1}(h)$, the level
of which is determined by the emotional information available at that
time. Only if $a_{1}(h)> \tau_{i}$, the agent expresses its
emotions, which consequently sets back $a_{i}(t)$ to zero, otherwise it
remains at this subcritical arousal level. On the other hand, if the
agent, because of some fluctuations, reaches the unstable negative
arousal level $a_{2}(h)$, the feedback of eqn (\ref{eq:arousal-square})
will further amplify the negative arousal to $-\infty$. This means that
the agent never again expresses its emotions and 'drops out'. If this
happens to many agents, a collective emotion cannot be sustained. Which
of the two cases is reached, crucially depends on the fluctuation
distribution. Looking at the example of Fig. \ref{fig:arousalSolutions},
we can verify that initial fluctuations (for $a=0$) should not reach the
level of 0.1, in order to prevent a 'dropout' of the agents.

The scenario looks different if, instead of a saturated dynamics with
$d_{2}<0$, we assume $d_{2}>0$, i.e. a superlinear growth in the
arousal. Inequality (\ref{eq:restr}) then defines the range of possible
values of $d_{2}$ to guarantee two real solutions. As one can verify,
there are real solutions already for very small values of $h$. In the
following, we only concentrate on the range of sufficiently large $h$ as
shown in Fig. \ref{fig:arousalSolutions}. Then, both real solutions are
negative and the one closer to zero is the unstable solution, whereas the
most negative solution is stable. For agents starting with a small
positive or negative arousal, $a>a_{1}(h)$, $a$ may further grow
independent on the value of $h$ until it reaches the threshold
$\tau_{i}$, at which the agent expresses its emotion, i.e. agents
do not remain at a subcritical arousal level forever. If the arousal is
set back to zero and because of fluctuations reaches negative values
$a<a_{1}(h)$, it will become more negative, but is always bound by the
negative stationary value $a_{2}(h)$. I.e., the agent never 'drops out'
entirely. Instead, even with a negative arousal, it can always get back
into an active regime dependent on the fluctuation
distribution. Consequently, the 'non-saturated' case defines the scenario
where we most likely expect the emergence of collective emotions, where
agents regularly express their emotions. However, such a scenario can
never be sustained in a purely deterministic dynamics. Instead,
spontaneous fluctuations are essential, and our analysis already tells us
the critical size of the fluctuations needed (determined by $A_{a}$). As
we can verify in Fig. \ref{fig:arousalSolutions}, this critical
fluctuation level depends on the total information $h$, which is not
unrealistic, because more (diverse) information is also associated with
more ambivalence.

\subsection{Expression of emotions}
\label{sec:expression}

So far, we have identified critical regimes both in the valence and in
the arousal dynamics, provided a given emotional information
$h$. However, as explained above, this information is only generated by
the agents above a critical arousal. Consequently, we need to ask what is
the minimal time lapse before an agent reaches the threshold $\tau$,
contributing to the emotional information. For simplicity, we take the
delay time in eqn. (\ref{s-i}) as $\Delta t = 0$ for all expressions. The
time lapse to reach the threshold is given by the dynamics of
eqn. \ref{eq:arousal-square}, which can be solved assuming a given value
of $h$:
\begin{equation}
  \label{eq:arousal-deq-sol}
  \int_{0}^{T} dt = T = \int_{0}^{\tau} \frac{d a}{h d_{2} a^{2} 
    + (h d_{1} - \gamma_{a})a + d_{0} h}
\end{equation}
This solution assumes that $h$ already exist, either because of an
external information, or because it is generated by other agents. Hence,
it is an adiabatic approximation of the full dynamics, which assumes
$\dot{h}=0$, this way describing the response of a single agent to the
existing (stationary) field. The solution of
eqn. (\ref{eq:arousal-deq-sol}) depends on whether the value of $R(h) =
4d_{2}d_{0} - (d_{1}- \gamma_a/h)^2$ is positive or negative.

\begin{figure}[h]
  \centering
  \includegraphics[width=0.48\textwidth]{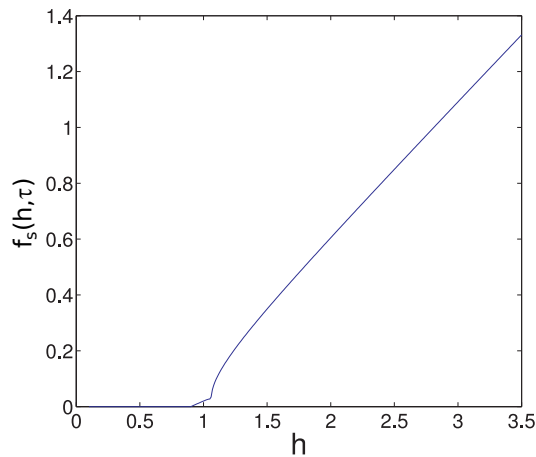}
  \caption{Frequency of emotional expression,
    $f_{s}(h,\tau)={1}/{T(h,\tau)}$ for $\tau=0.5$ and the parameters
    \ref{eq:parameters}.  Below a critical value of $h$, an agent with
    threshold $\tau$ would not express its emotions, but above the
    frequency of expression grows with the field $h$.}
  \label{fig:frequency}
\end{figure}

Following the discussion in the previous section, we now have to consider
two different regimes for arousal dynamics, the saturated one ($d_{2}<0$)
and the superlinear ($d_{2}>0$). In the saturated regime, always $R(h)<0$
and the solution is given by:
\begin{equation}
  \label{eq:interexpression-time-tanh}
  T(h,\tau) = \frac{2}{h \sqrt{-R(h)}} \arctanh \left (  \frac{\sqrt{-R(h)}}{2
      d_0/{\tau} + d_1 - \gamma_a/h } \right )
\end{equation}
For $d_{2}>0$ we can have both $R(h)<0$ and $R(h)>0$ dependent on the
choice of the other parameters. For $R(h)>0$, the solution of
eqn. (\ref{eq:arousal-deq-sol}) is given by:
\begin{equation}
  \label{eq:interexpression-time-tan}
  T(h,\tau) = \frac{2}{h \sqrt{R(h)}} \arctan \left ( \frac{\sqrt{R(h)}}{2 d_0/{\tau} 
      + d_1 - \gamma_a/h } \right )
\end{equation}
In the superlinear regime, we expect that the agent is likely to express
its emotions more than once (dependent on the fluctuations). In this
case, $T(h,\tau)$ gives the (idealized) periodicity of expressing
the emotion, i.e. the time after which the agent on average reaches the
threshold $\tau$ again, after it was set back to zero when
expressing the emotion last time. Fig. \ref{fig:frequency} shows the
frequency $f_{s}(h,\tau)=1/T(h,\tau)$ at which an agent
expresses its emotions, dependent on the (quasistationary) value of
$h$. We note that there is a nonmonotonous increase, i.e. below a
critical value of $h=h^{\star}$ the frequency is zero, i.e. we do not
expect a collective emotional state where agents more than once express
their emotions, whereas for $h>h^{\star}$, agents may regularly
contribute emotional information, which means a collective emotion is
sustained.

\begin{figure}[h]  
  \centering
  \includegraphics[width=0.48\textwidth]{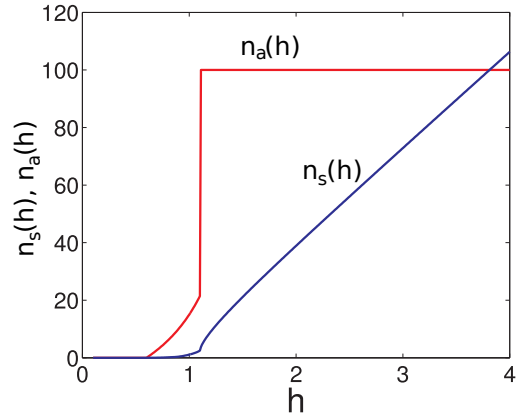}
  \caption{ $n_{a}(h)$ (red) and $n_{s}(h)$ (blue) versus field $h$ for
    the parameter set \ref{eq:parameters}.  Above a critical level of
    information $h$, agents start to participate in the
    conversation. Their number $n_{a}(h)$ grows very fast until the whole
    community is involved.}
  \label{fig:nAgents}
\end{figure}

Based on the frequency of expression, we are able to calculate the
average number of expressions per time intervall, $n_{s}$, as defined in
eqn. (\ref{eq:ns-average}).  Assuming $N$ agents with a threshold
distribution $P(\tau)$, the number of agents with a given threshold
$\tau$ is $N(\tau)=NP(\tau)$, whereas the frequency $f_{s}(h,\tau)$
defines how often such agents reach an arousal above the threshold,
forcing them to express their emotions. Assuming a uniform threshold
distribution for simplicity, we can calculate for a given $h$:
\begin{equation}
  n_{s} = N \int f_{s}(h,\tau) P(\tau) d\tau = 
  \frac{N}{\tau_{\mathrm{max}}-\tau_{\mathrm{min}}} 
  \int^{\tau_{\mathrm{max}}}_{\tau_{\mathrm{min}}} \frac{d\tau}{T(h,\tau)}
  \label{eq:ns-tau}
\end{equation}
where $T(h,\tau)$ is given by eqs. (\ref{eq:interexpression-time-tanh}),
(\ref{eq:interexpression-time-tan}).  $n_{s}$ is plotted in
Fig. \ref{fig:nAgents}. Above a critical value of the field $h^{\star}$,
the number of expressions per time interval increases monotonously, with
a noticable knee at the point where all agents become
involved. Obviously, for lower values of $h$ not all agents reach an
arousal above the threshold, which prevents them from expressing their
emotions. But at a characteristic value $\hat{h}$, the field is large
enough to bring all their arousals above the threshold. 

Similar to eqn. (\ref{eq:ns}), we can also calculate the number of agents
expressing their emotions at any given time $t$ as:
\begin{equation}
  \label{eq:na}
  N_{a}(t)=\sum_{i} 1-\Theta\left[-f_{s}(h,\tau)\right]
\end{equation}
Again $\Theta[x]$ is one only if $x\geq 0$, i.e. for agents with
frequency zero the Heavyside function $\Theta[-x]$ returns one. The
\emph{average} number of agents expressing their emotions per time
interval is then, similar to
eqs. (\ref{eq:ns-average}),(\ref{eq:ns-tau}):
\begin{equation}
  \label{eq:na-average}
  n_{a}=\frac{1}{t_{\mathrm{end}}}\int_{0}^{t_{\mathrm{end}}}N_{a}(t) dt
  =  N \int \left( 1- \Theta\left[-f_{s}(h,\tau)\right]\right) P(\tau)
  d\tau 
\end{equation}
which can be calculated similar to eqn. (\ref{eq:ns-tau}). $n_{a}$ is
plotted in Fig. \ref{fig:nAgents} as well, and one clearly identifies the
critical $\hat{h}$, to involve all agents.

\section{Computer simulations of collective emotions}
\label{sec:simulation}

Based on the analytical insights obtained, we eventually present the
results of agent-based computer simulations. This means that we've
implemented the individual dynamics given by the stochastic
eqs. (\ref{eq:v-basic}), (\ref{eq:a-basic}) of $N$ agents with a
heterogeneous threshold distribution $P(\tau)$. The latter one is
important as the process of forming a collective emotion needs generating
emotional information, $h(t)$. There could be two possibilities to start this
process: (i) an external trigger, expressed by $I_{\pm}(t)$ in
eqn. (\ref{h}), (ii) initial fluctuations in the arousal which have to be
large enough to push some of the agents above the threshold. Very similar
to the model of social activation \citep{granovetter}, it then depends on
the distribution of thresholds and the feedback dynamics whether more
agents become involved. For our simulations, we have chosen the
parameters for the valence and the arousal dynamics, $b_{k}$, $d_{k}$ in
such a way that a supercritical feedback between the emotional
information generated and the activity of the agents is
guarenteed. Specifically, we have chosen: 
\begin{eqnarray}
&&  \gamma_v = 0.5,\; A_v = 0.3,\; b_{1}=1,\: b_{3}=-1\;\gamma_h = 0.7,\; A_a = 0.3,\;
  \gamma_{a} = 0.9, \nonumber \\
&& d_0 = 0.05,\; d_1 = 0.5,\; \tau_{\mathrm{min}} = 0.1,\; \tau_{\mathrm{max}} = 1.1,\; N = 100,\; s =
  0.1,\; h_0= 0
\label{eq:parameters}
\end{eqnarray}
That means that, thanks to our analytical efforts, we are likely to
expect a collective emotion where most agents express their emotions at
least once. Our main focus is therefore on the two different scenarios
expressed by the parameter $d_{2}\lessgtr 0$, which result from the
saturated or the superlinear feedback of emotional information on the
arousal dynamics.

\begin{figure}[htbp]
  \centering
    \includegraphics[width=0.48\textwidth]{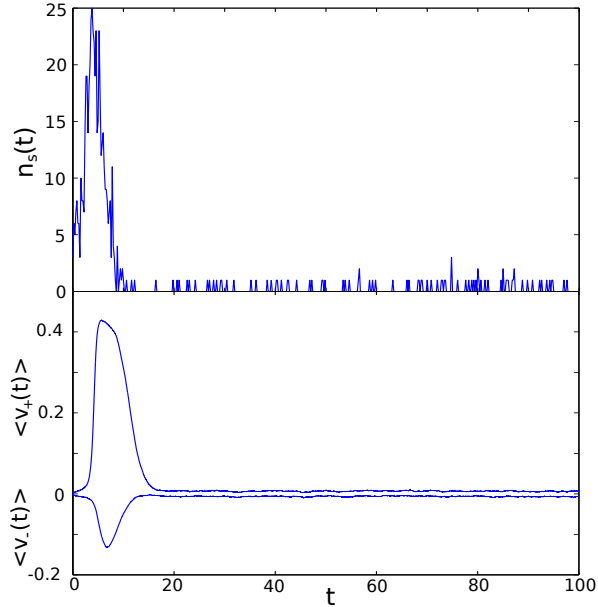} \hfill
    \caption{Agent expressions (top) binned with $\delta t=0.2$ and
      average positive and negative valences (bottom) and for a
      simulation with parameters \ref{eq:parameters} and $d_2 = -0.1$.  A
      collective emotional state appears and disappears after some time,
      but never reappears again.}
  \label{fig:bad}
\end{figure}

In the saturated case, $d_{2}<0$, we expect that a collective emotion may
appear, but not be sustained because agents have a tendency to 'drop
out'. This scenario is illustrated in Fig. \ref{fig:bad}. Calculating the
average number of expressions per time interval, $n_{s}(t)$, eqn.
(\ref{eq:ns-average}), we observe an initial burst of activity in the
beginning, i.e. many agents contribute their emotional information, which
then fades out, only keeping a random level of activity. That means, we
observe indeed the emergence of a collective emotion, but this is not
sustained because of the assumed saturation. This is also confirmed in
Fig. \ref{fig:bad}, which shows the averaged positive and negative
valences of agents. We observe the emergence of a polarized state, where
agents with strong positive and mildly negative emotions coexist, i.e. a
bimodal valence distribution appears and remains for a while, before it
disappears completely because agents 'dropped out'. Consequently, the
saturated regime allows the emergence of a collective emotion, but it is
restricted to appear once and never again.

\begin{figure}[htbp]
  \centering
    \includegraphics[width=0.48\textwidth]{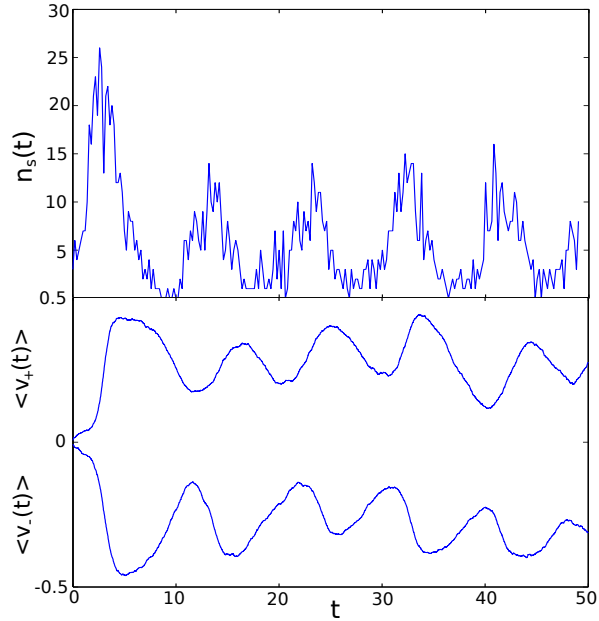} \hfill
    \caption{Agent expressions (top) binned with $\delta t=0.2$ and
      average positive and negative valences (bottom) and for a
      simulation with parameters \ref{eq:parameters} and $d_2 = 0.5$.  A
      collective emotional state appears, fades out and reappears again.
    }
  \label{fig:new}
\end{figure}

In the superlinear case, $d_{2}>0$, we expect the emergence of collective
emotions more than once, i.e. they can fade out and be reestablished
again. We consider this the more realistic scenario for applications to
internet users, where the up and downs of collective emotions are indeed
observed. Fig. \ref{fig:new} illustrates this scenario in a way
comparable to Fig. \ref{fig:bad}. Here, we see waves of activity
indicated by the number of emotional expressions per time interval. The
respective averaged positive and negative valences also reflect these
waves, i.e. we observe more or less polarized states dependent on the
activity.
 
\begin{figure}[htbp]  
  \centering
  \includegraphics[width=0.48\textwidth]{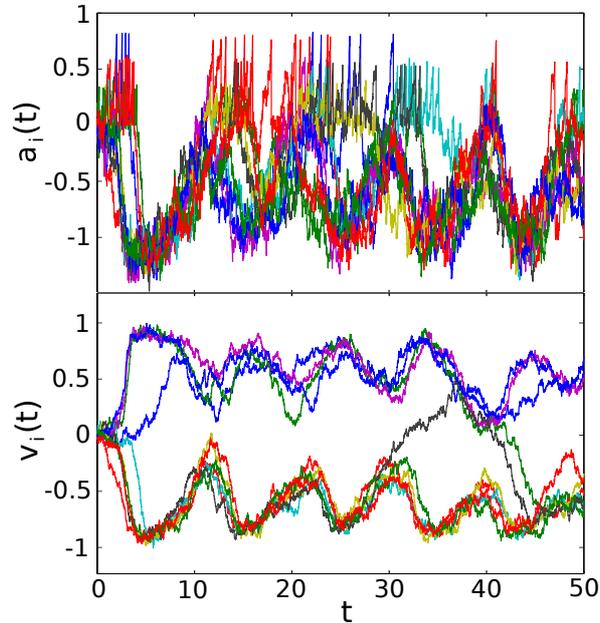}\hfill
    \caption{Sample trajectories of the arousal (top) and the valence
      (bottom) of ten agents in a simulation with parameters as in figure
    \ref{fig:new}.}
\label{fig:sample}
\end{figure}

Consequently, the collective emotions not only emerge once, but are also
sustained over a long period. The reason for this was already explained
in Sect. \ref{sec:arousal-sim}. If agents have expressed their emotions
and fall into a 'careless' state characterized by negative arousal, no
new emotional information is produced. This in turn lowers the field $h$,
which determines the stationary value of the negative arousal at which
agents 'rest'. The lower the field, the larger the stationary arousal,
which eventually allows the fluctuations to push agents back into an
active regime of $a(t)>a_{1}(h)$. To illustrate this, we have plotted the
arousal of ten randomly chosen agents in Fig. \ref{fig:sample}. The
typical oscilatory behavior can be clearly seen. If the field is
initially low, most likely  $a(t)>a_{1}(h)$, i.e. agents arousal is
increased until they express their emotions. This generates a high
field. If agents arousal is set back to zero at high $h$, $a_{1}(h)$ is
almost zero (as can be verified in Fig. \ref{fig:arousalSolutions}),
which means that most agents reach the stable stationary level of
negative $a_{2}(h)$, at which they remain until $h$ is lowered again. 

The corresponding dynamics of the valence for the randomly chosen agents
is also shown in Fig. \ref{fig:sample}. One can notice a quite
synchronized change of the emotions, which is not surprising as the
dynamics mainly depends on the value of $h$, which is the same for all
agents and all other parameters are kept constant. We can, of course,
consider more heterogeneous parameters for the agents, to allow for more
diversity. We wish to emphasize that agents do not always have the same
emotion over time, some of the sample trajectories clearly show that
agents switch from positive to negative emotions and vice versa.

To conclude, all simulations are consistent with the analytical results
derived in the previous sections. Based on these results, we may be able
to derive hypotheses about the behavior of emotional agents, which can be
tested e.g. in psychological experiments.

\section{Conclusion}
\label{sec:conclusion}

The aim of our paper is to provide a general framework for studying the
emergence of collective emotions in online communities. I.e. we are not
particularly interested in the most complete description of individual
emotions, but rather in an approach that allows to generate testable
hypotheses about the conditions under which a particular collective
dynamics can be observed. Nevertheless, we refrain from using ad hoc
assumptions about the dynamics of individual agents which have been used
in 'sociophysics' models of opinion dynamics, etc. Instead, our starting
point is indeed a psychological theory of how individual emotional states
should be described. Hence, our agent model is based on on
\emph{psychological variables} such as arousal $a$ and valence $v$. 

As the second important ingredient of our general framework, we
explicitely address the \emph{communication} between agents. This is very
important to model online communities, where agents do not have a direct,
face-to-face communication, but an indirect, time delayed communication,
which is mediated by a medium. The latter stores the information
expressed by the agents (mostly in terms of writings) and allows all
agents to get access to this information at the same or a different
times. Consequently, this medium provides a mean-field coupling between
all agents, which is reflected by the so-called communication field $h$
in our framework. By explicitly modeling the dynamics of the information
stored, we consider the decentralized generation of emotional information
of different types ($h_{+}$, $h_{-}$), at different times, the 'aging' of
emotional information (i.e. the decrease in impact of older information)
and the distribution of information among agents, for which also other
than mean-field assumptions can be used. 

As the third ingredient, we eventually model the impact of the emotional
information on agents dependent on their emotional states. Here we assume
a very general \emph{nonlinear feedback} between the available
information $h$ and the individual valence $v$ and arousal $a$. This
allows to derive different hypotheses about the impact, which can be
tested e.g. in psychological experiments. On the other hand, if such
insight should become available to us, we are able to cope with these
findings and to check their consequences on the emergence of collective
emotions

It is a strength of our general framework that it allows an analytical
treatment, to estimate the range of parameters under which the emergence
of a collective emotion can be expected. In particular, we are able to
specify the conditions for (a) a polarized collective emotional state
(bimodal valence distribution), and (b) scenarios where agents express
their emotions either once or consecutively (saturated vs superlinear
impact on the arousal dynamics). Such findings are important in order to
later calibrate the model parameters against empirical data describing
these different regimes.

Hence, our modeling approach offers a link to both psychological
experiments with individuals, testing the hypotheses about the impact of
emotional information, and to data analysis of emotional debates among
internet users, determining model parameters for different scenarios.

Eventually, the general framework provided here is extensible and
flexible enough to encompass different situations where collective
emotions emerge. This is thanks to the distinction of the two different
layers already depicted in Fig. \ref{fig:general-schema}: the internal
layer describing the agent and its emotional states and the external
layer describing the communication process of expressing emotional
information. If we, for example, want to apply this framework to emotions
expressed in product reviews as e.g. analysed in  \citep{Lorenz2009}, we have to
consider that agents usually review a product only once, i.e. the
saturated scenario for the arousal is more appropriate here. The dynamics
for the valence, describing the emotional content, has to consider that
the feeling of the agent also depend on the product quality $q$ and the
user preference $u_{i}$, i.e. $v_{i}\propto |u_{i}-q|$. Already such
extensions are able to reproduce the distribution of the emotional
content in reviews to a very remarkable degree \citep{David}. Considering
marketing campaigns in terms of external information versus a sole
word-of-mouth spreading of emotional information also allows to capture
different observed scenarios in generating emotional ratings
\citep{David}.

In addition to these promising applications, there are other models which
can be recasted in our general framework. For example, \citep{rank2010} or
\citep{Czaplicka2010} have proposed agent-based models, where the information
field does not feed back on the agent's arousal, which is assumed as an
exogeneous constant probability of action. Consequently, instead of
modeling the internal arousal dynamics explicitely as proposed in the
general framework, these two models rather focus on the feedback between
the expressed valence of different agents, which makes them special cases
of the general framework.

To conclude, with our agent-based model we have provided a general
framework to understand and to predict the emergence of collective
emotions based on the interaction of agents with individual emotional
states. As the framework is very tractable both in terms of mathematical
analysis and computer simulations, we are now working on applying it to
emotional debates observed in different online communities, such as in
blogs, newsgroups, or discussion fora.

\section*{Acknowledgement}

The research leading to these results has received funding from the
European Community's Seventh Framework Programme FP7-ICT-2008-3 under
grant agreement no 231323 (CYBEREMOTIONS).

\bibliographystyle{sg-bibstyle} \bibliography{paper1}

\begin{thebibliography}{30}
\expandafter\ifx\csname natexlab\endcsname\relax\def\natexlab#1{#1}\fi
\expandafter\ifx\csname url\endcsname\relax
  \def\url#1{\texttt{#1}}\fi
\expandafter\ifx\csname urlprefix\endcsname\relax\def\urlprefix{URL }\fi
\expandafter\ifx\csname selectlanguage\endcsname\relax
  \def\selectlanguage#1{\relax}\fi

\bibitem[{Adams and Roscigno(2005)}]{Adams2005}
Adams, J.; Roscigno, V.~J. (2005).
\newblock White Supremacists, Oppositional Culture and the World Wide Web.
\newblock \emph{Social Forces} \textbf{84(2)}, 759.

\bibitem[{Bar-Tal \emph{et~al.}(2007)Bar-Tal, Halperin and
  de~Rivera}]{Bartal2007}
Bar-Tal, D.; Halperin, E.; de~Rivera, J. (2007).
\newblock Collective Emotions in Conflict Situations: Societal Implications.
\newblock \emph{Journal of Social Issues} \textbf{63(2)}, 441.

\bibitem[{Barrett and Russell(1999)}]{FeldmanBarret1999}
Barrett, L.~F.; Russell, J.~A. (1999).
\newblock The Structure of Current Affect: Controversies and Emerging
  Consensus.
\newblock \emph{Current Directions in Psychological Science} \textbf{8(1)}.

\bibitem[{Bradley(2009)}]{bradley2009}
Bradley, M.~M. (2009).
\newblock Natural selective attention: Orienting and emotion.
\newblock \emph{Psychophysiology} \textbf{46(1)}, 1--11.

\bibitem[{Chmiel and Holyst(2010)}]{Chmiel2010}
Chmiel, A.; Holyst, J.~A. (2010).
\newblock Flow of emotional messages in artificial social networks.
\newblock \emph{International Journal of Modern Physics C} \textbf{21(5)},
  593--602.

\bibitem[{Czaplicka \emph{et~al.}(2010)Czaplicka, Chmiel and
  Holyst}]{Czaplicka2010}
Czaplicka, A.; Chmiel, A.; Holyst, J.~A. (2010).
\newblock Emotional Agents at the Square Lattice.
\newblock \emph{Acta Physica Polonica A} \textbf{117(4)}, 688--694.

\bibitem[{DiMaggio \emph{et~al.}(2001)DiMaggio, Hargittai, Neuman and
  Robinson}]{DiMaggio2001}
DiMaggio, P.; Hargittai, E.; Neuman, W.~R.; Robinson, J.~P. (2001).
\newblock Social Implications of the Internet.
\newblock \emph{Annual Review of Sociology} \textbf{27}, 307.

\bibitem[{Dodds and Danforth(2009)}]{Dodds2009}
Dodds, P.~S.; Danforth, C.~M. (2009).
\newblock Measuring the Happiness of Large-Scale Written Expression: Songs,
  Blogs, and Presidents.
\newblock \emph{Journal of Happiness Studies} .

\bibitem[{Flache(2004)}]{Flache2004}
Flache, A. (2004).
\newblock How May Virtual Communication Shape Cooperation in a Work Team?
\newblock \emph{Analyse \& Kritik} \textbf{26}, 258.

\bibitem[{Garcia and Schweitzer()}]{David}
Garcia, D.; Schweitzer, F. ().
\newblock Emotions in product reviews -- Empirics and models.
\newblock (to be submitted).

\bibitem[{Gianotti \emph{et~al.}(2008)Gianotti, Faber, Schuler, Pascual-Marqui,
  Kochi and Lehmann}]{gianotti2008first}
Gianotti, L.; Faber, P.; Schuler, M.; Pascual-Marqui, R.; Kochi, K.; Lehmann,
  D. (2008).
\newblock {First valence, then arousal: the temporal dynamics of brain electric
  activity evoked by emotional stimuli}.
\newblock \emph{Brain Topography} \textbf{20(3)}, 143--156.

\bibitem[{Granovetter(1978)}]{granovetter}
Granovetter, M. (1978).
\newblock {Threshold Models of Collective Behavior}.
\newblock \emph{American Journal of Sociology} \textbf{83(6)}, 1420.

\bibitem[{Gratch and Marsella(2004)}]{gratch04}
Gratch, J.; Marsella, S. (2004).
\newblock A domain-independent framework for modelin emotion.
\newblock \emph{Cognitive Systems Research} \textbf{5(4)}, 269--306.

\bibitem[{Gratch \emph{et~al.}(2009)Gratch, Marsella, Wang and
  Stankovic}]{gratch09}
Gratch, J.; Marsella, S.; Wang, N.; Stankovic, B. (2009).
\newblock Assessing the validity of appraisal-based models of emotion.
\newblock \emph{International Conference on Affective Computing and Intelligent
  Interaction} .

\bibitem[{Jones and Troen(2007)}]{Jones2007}
Jones, C.~M.; Troen, T. (2007).
\newblock Biometric Valence and Arousal Recognition.
\newblock In: \emph{OZCHI '07: Proceedings of the 19th Australasian conference
  on Computer-Human Interaction}, ACM.
\newblock ISBN 978-1-59593-872-5, p. 191.

\bibitem[{Lorenz(2009)}]{Lorenz2009}
Lorenz, J. (2009).
\newblock Universality in movie rating distributions.
\newblock \emph{European Physical Journal B} \textbf{71(2)}, 251--258.

\bibitem[{Lorenz \emph{et~al.}(2009)Lorenz, Battiston and Schweitzer}]{lorenz}
Lorenz, J.; Battiston, S.; Schweitzer, F. (2009).
\newblock Systemic Risk in a Unifying Framework for Cascading Processes on
  Networks.
\newblock \emph{European Physical Journal B} \textbf{71(4)}, 441--460.

\bibitem[{de~Melo \emph{et~al.}(2009)de~Melo, Zheng and Gratch}]{melo09}
de~Melo, C.~M.; Zheng, L.; Gratch, J. (2009).
\newblock Expression of Moral Emotions in Cooperating Agents.
\newblock \emph{Intelligent Virtual Agents} .

\bibitem[{Mitrovi\'c and Tadi\'c(2010)}]{Mitrovic2009}
Mitrovi\'c, M.; Tadi\'c, B. (2010).
\newblock Bloggers behavior and emergent communities in Blog space.
\newblock \emph{European Physical Journal B} \textbf{73(2)}, 293--301.

\bibitem[{Prabowo and Thelwall(2009)}]{Prabowo2009}
Prabowo, R.; Thelwall, M. (2009).
\newblock Sentiment analysis: A combined approach.
\newblock \emph{Journal of Informetrics} \textbf{3(2)}, 143 -- 157.
\newblock ISSN 1751-1577.

\bibitem[{Rank(2010)}]{rank2010}
Rank, S. (2010).
\newblock Docking Agent-based Simulation of Collective Emotion to
  Equation-based Models and Interactive Agents.
\newblock In: \emph{Proceedings of Agent-Directed Simulation Symposium, 2010
  Spring Simulation Conference.} pp. 82--89.

\bibitem[{Russell(1980)}]{Russell1980}
Russell, J.~A. (1980).
\newblock A Circumplex Model of Affect.
\newblock \emph{Journal of Personality and Social Psychology} \textbf{39(6)},
  1161.

\bibitem[{Sassenberg and Boos(2003)}]{Sassenberg2003}
Sassenberg, K.; Boos, M. (2003).
\newblock Attitude Change in Computer-Mediated Communication: Effects of
  Anonymity and Category Norms.
\newblock \emph{Group Processes \& Intergroup Relations} \textbf{6(4)}, 405.

\bibitem[{Scherer \emph{et~al.}(2001)Scherer, Schorr and Johnstone}]{scherer01}
Scherer, K.~R.; Schorr, A.; Johnstone, T. (eds.) (2001).
\newblock \emph{Appraisal processes in emotion: theory, methods, research}.
\newblock Oxford University Press.

\bibitem[{Scherer \emph{et~al.}(2004)Scherer, Wranik, Sangsue, Tran and
  Scherer}]{Scherer2004}
Scherer, K.~R.; Wranik, T.; Sangsue, J.; Tran, V.; Scherer, U. (2004).
\newblock Emotions in everyday life: probability of occurrence, risk factors,
  appraisal and reaction patterns.
\newblock \emph{Trends and developments: Research on Emotions} \textbf{43(4)},
  499.

\bibitem[{Schweitzer(2003)}]{schweitzer03c}
Schweitzer, F. (2003).
\newblock \emph{Brownian Agents and Active Particles. Collective Dynamics in
  the Natural and Social Sciences}.
\newblock Berlin: Springer.
\newblock With a foreword by J. Doyne Farmer.

\bibitem[{Schweitzer and Holyst(2000)}]{schweitzer00b}
Schweitzer, F.; Holyst, J. (2000).
\newblock Modelling collective opinion formation by means of active {B}rownian
  particles.
\newblock \emph{European Physical Journal B} \textbf{15(4)}, 723 -- 732.

\bibitem[{Sobkowicz and Sobkowicz(2010)}]{Sobkowicz2010}
Sobkowicz, P.; Sobkowicz, A. (2010).
\newblock Dynamics of hate based Internet user networks.
\newblock \emph{European Physical Journal B} , 633.

\bibitem[{Thelwall \emph{et~al.}(2010)Thelwall, Wilkinson and
  Uppal}]{Thelwall2009}
Thelwall, M.; Wilkinson, D.; Uppal, S. (2010).
\newblock Data mining emotion in social network communication: Gender
  differences in MySpace.
\newblock \emph{Journal of the American Society for Information Science and
  Technology} \textbf{61}, 190--199.

\bibitem[{Yik \emph{et~al.}(1999)Yik, Russell and Barret}]{Yik1999}
Yik, M.; Russell, J.; Barret, L.~F. (1999).
\newblock Structure of Self-reported Affect: Integration and Beyond.
\newblock \emph{Journal of Personal and Social Psychology} \textbf{77(3)}, 600.

\end{thebibliography}

\end{document}